\documentclass[final,journal]{IEEEtran}
\usepackage{amsmath,amsthm}

\theoremstyle{plain}

\theoremstyle{definition}

\usepackage{graphicx}  
\usepackage{subcaption}  
\theoremstyle{remark}

\usepackage{graphicx}
\usepackage{amssymb} 
\usepackage{enumitem}
\usepackage{booktabs}
\usepackage{listings}
\usepackage{mathrsfs}
\usepackage{textcomp}
\usepackage{epstopdf}
\usepackage{pdfpages}
\usepackage{cite}
\usepackage[normalem]{ulem}
\usepackage{algorithm}
\usepackage{algorithmic}
\usepackage{array}
\usepackage{arydshln}
\usepackage{xcolor}

\begin{document}

\title{Sustainable Task Offloading in Secure UAV-assisted Smart Farm Networks: A Multi-Agent DRL with Action Mask Approach 
} 

\author{Tingnan Bao, \IEEEmembership{Senior Member, IEEE}, Aisha Syed, William Sean Kennedy, Melike Erol-Kantarci, \IEEEmembership{Senior Member, IEEE} 
\thanks{T. Bao and M. Erol-Kantarci are with the School of Electrical Engineering and Computer Science, University of Ottawa, Ottawa, ON, Canada (emails: tbao, melike.erolkantarci@uottawa.ca)}
\thanks{A. Syed and W.-S. Kennedy are with Nokia Bell Labs (emails: aisha.syed, william.kennedy@nokia-bell-labs.com)} }

\maketitle

\begin{abstract}
The integration of unmanned aerial vehicles (UAVs) with mobile edge computing (MEC) and Internet of Things (IoT) technology in smart farms is pivotal for efficient resource management and enhanced agricultural productivity sustainably. This paper addresses the critical need for optimizing task offloading in secure UAV-assisted smart farm networks, aiming to reduce total delay and energy consumption while maintaining robust security in data communications. We propose a multi-agent deep reinforcement learning (DRL)-based approach using a deep double Q-network (DDQN) with an action mask (AM), designed to manage task offloading dynamically and efficiently. The simulation results demonstrate the superior performance of our method in managing task offloading, highlighting significant improvements in operational efficiency by reducing delay and energy consumption. This aligns with the goal of developing sustainable and energy-efficient solutions for next-generation network infrastructures, making our approach an advanced solution for achieving both performance and sustainability in smart farming applications.
\end{abstract}


\section{Introduction}
The emergence of Internet of Things (IoT) technology has revolutionized numerous industries, including agriculture. Smart agriculture leverages IoT to enhance resource management and improve crop productivity through data-driven decision-making \cite{chettri2019comprehensive,elijah2018overview,peng2021analysis}. This innovation significantly increases the efficiency and sustainability of agricultural practices. Task offloading becomes a key component of this technology integration, where computing tasks generated by IoT devices are systematically offloaded to more powerful processors \cite{pamuklu2023heterogeneous,kumar2022iot,zhu2019novel}. This strategy is critical to overcome the inherent limitations of IoT device processing capabilities, which may otherwise hinder real-time data processing and timely decision-making. By offloading these tasks, smart agricultural operations can maintain continuous and effective monitoring and management of the agricultural environment, ensuring optimal resource utilization and increased productivity. The integration of IoT in agriculture and the specific role of task offloading has been widely documented in the literature, with many studies highlighting its benefits and applications in enhancing the efficiency and sustainability of modern agricultural practices\cite{dhanaraju2022smart}.

Unmanned aerial vehicles (UAVs) have become crucial in advancing IoT frameworks in agriculture, serving as dynamic nodes that significantly enhance the offloading process through the establishment of high-quality line-of-sight (LoS) connections \cite{li2018uav}. However, the operational capacity of UAVs is often constrained by their limited battery life and computing power, which may not support prolonged or complex task processing \cite{mozaffari2017mobile}. To address these limitations, mobile edge computing (MEC) servers play an essential role by handling tasks that demand substantial computational power and energy, thus complementing the capabilities of UAVs within these networks. The integration of MEC servers is vital for overcoming the computational limitations of UAVs, enabling more efficient processing and energy management in UAV-assisted networks. The literature extensively explores various task offloading strategies within these UAV-assisted networks, emphasizing the importance of optimizing energy consumption and improving performance. \cite{nguyen2022reinforcement, han2020rate, pamuklu2022iot, ebrahim2022deep}. In \cite{nguyen2022reinforcement}, the authors investigated task offloading challenges, focusing on maximizing UAV hover times in UAV-assisted systems. The authors in \cite{han2020rate} explored enhancements to computation capacity performance in similar networks. Innovative risk-sensitive and delay-sensitive task offloading strategies have been introduced in \cite{pamuklu2022iot} that optimize the hovering times of UAVs, effectively balancing operational demands with energy conservation in UAV-assisted networks. In \cite{ebrahim2022deep}, the authors discussed strategic task offloading decisions that strike a balance between rapid data processing and minimal energy consumption, thereby maximizing overall efficiency in IoT networks.

While the integration of MEC servers significantly enhances the computational capabilities and efficiency of UAV-assisted networks, it concurrently results in a substantial increase in data transmissions, which consequently become more susceptible to eavesdropping. To mitigate this vulnerability, physical layer security (PLS) provides a robust solution for securing UAV-assisted networks by leveraging the unique characteristics of wireless channels \cite{chen2022secure, chen2023secure, bai2019energy, lu2023resource}. In \cite{chen2022secure}, the authors developed a secure offloading scheme using the block coordinate descent (BCD) method and successive convex approximation (SCA) technique to maximize the minimum average secrecy capacity in UAV-assisted MEC networks. Similarly, \cite{chen2023secure} introduced an efficient BCD-based iterative algorithm designed to maximize the minimum average utility of an MEC-based UAV-unmanned ground vehicle (UGV) collaborative framework, focusing on secure task offloading for rural area surveillance. The study in \cite{bai2019energy} investigated an energy-efficient computation offloading scheme in UAV-assisted MEC networks, considering both passive and active eavesdropper. Moreover, \cite{lu2023resource} proposed an efficient secure communication scheme by jointly optimizing resource allocation and trajectory planning in UAV-relay-assisted secure maritime MEC environments, with the presence of a flying eavesdropper.

However, most of the aforementioned works rely on multiple iterative methods to solve the formulated problems, leading to significant computational complexity. To address this issue, machine learning (ML) techniques have been utilized to effectively manage high-dimensional resource management challenges \cite{jiang2019deep,wang2021deep,seid2021multi,zhao2022multi,wu2024joint,li2020onboard,zhang2021task,nguyen2022deep}. For instance, in \cite{jiang2019deep}, the authors proposed three deep learning (DL) algorithms aimed at minimizing the total energy consumption in UAV-assisted MEC networks. Furthermore, the authors  employed deep reinforcement learning (DRL) to jointly optimize resource allocation and UAV positioning in \cite{wang2021deep} . Additionally, a multi-agent DRL scheme was introduced in \cite{seid2021multi} and further explored in \cite{zhao2022multi} to minimize the overall network computation cost in UAV-assisted networks.  Moreover, the authors investigated and solved the long-term optimization problem of joint task offloading and resource allocation in a multi-UAV multi-server MEC network using a proposed DRL scheme \cite{wu2024joint}. However, few of the existing works in UAV-assisted MEC networks have considered the priority of task offloading. As a result, all tasks are equally likely to be offloaded to servers, receiving similar computing resources regardless of their urgency. This uniform approach can lead to significant inefficiencies, particularly for tasks with strict latency requirements that may not be completed within their designated maximum allowable latency. 

To the best of our knowledge, a multi-agent DRL scheme for priority-driven task offloading in the secure UAV-assisted smart farm network has not yet been investigated. Therefore, this paper introduces a novel multi-agent DRL scheme that integrates a deep double Q-network (DDQN) with an action mask (AM) to efficiently manage task offloading in secure UAV-assisted smart farm networks. Our proposed DDQN with AM is designed to minimize the overall system cost by reducing total delay and energy consumption, while simultaneously ensuring secure communications. This is achieved by prioritizing tasks based on their urgency and importance. The scheme is under several constraints, including delay limitations for both local and edge processing, required secrecy rates, energy usage for transmission and processing activities, and resource constraints on UAVs for priority-driven task offloading. Through this approach, we aim to enhance the operational efficiency and security of smart farming operations, ensuring timely and reliable agricultural data processing and decision-making. The main contributions are summarized as follows:
\begin{itemize}
\item We have formulated the priority-driven task offloading problem within a secure UAV-assisted smart farm network. This problem is characterized as a mixed-integer non-linear program (MINLP) and recognized as NP-hard.This highlights the significant computational complexity and challenges in achieving feasible solutions, underscoring the need for efficient, scalable approaches to task offloading that align with sustainability and energy efficiency goals.

\item To address the NP-hard nature of the problem, we propose a novel multi-agent DRL based on AM framework. This framework is specifically designed to efficiently manage task offloading in secure UAV-assisted smart farm networks, thereby optimizing resource allocation and operational responses in real-time.
    
\item We implement a training process coupled with a distributed execution stage in our proposed DDQN with AM scheme. This dual-phase approach ensures that each UAV not only contributes to the overarching network goals but also effectively navigates and responds to its immediate environment, thereby enhancing adaptability, operational efficiency, and sustainability in smart agricultural practices.

\item Simulation results validate the effectiveness of our proposed DDQN with AM framework, demonstrating superior performance in terms of convergence, delay reduction, and energy conservation to baseline models. These results highlight the practical benefits and the enhanced decision-making capabilities of our approach in managing complex task offloading scenarios in smart farming applications, thereby contributing to the development of sustainable, energy-efficient network infrastructures.
\end{itemize}

The remainder of the paper is outlined as follows. Section II introduces the system model and problem formulation of secure UAV-assisted communication. Section III provides a detailed description of the proposed approach. Section IV presents the simulation results, and finally, Section V concludes of this paper.

\section{System model and Problem Formulation}

\begin{figure}[t!]
\centering
\includegraphics[height=6.8cm,width=7.8cm]{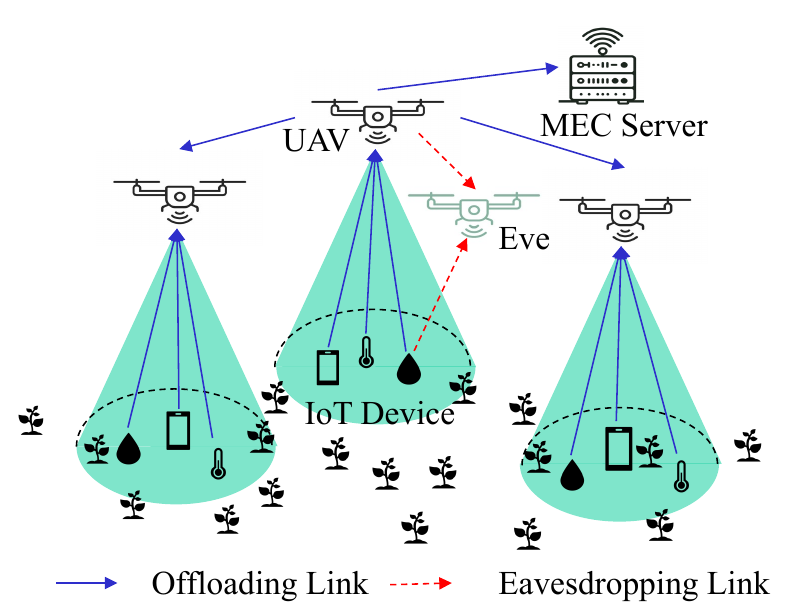}
\caption{Illustration of a secure UAV-assisted smart farm network.}
\label{fig_sys_mdl}
\end{figure}

We explore a secure UAV-assisted communication system deployed in a smart farm, as illustrated in Fig. \ref{fig_sys_mdl}. The system consists of $N$ IoT devices, $M$ legitimate UAVs, and a single MEC server, termed as $s$, all operating under the presence of an eavesdropping UAV, known as Eve $E$. Within this setup, every IoT device establishes communication with a its own UAV. Owing to the limited computational capacities of IoT devices, each device is equipped with a sensor tasked with collecting data pertinent to a specific activity out of $K$ possible types. This information, harvested from the environment, is subsequently transmitted to the assigned UAV. However, due to the finite computing and power resources of each legitimate UAV, it faces the decision of either processing tasks locally, offloading them to the MEC server, or another legitimate UAV.

We define the task set, the IoT device set, and the legitimate UAV set as $\mathbf{K} = \{1,\cdots, K \} $, $\mathbf{N} = \{1,\cdots, N \} $, and $\mathbf{M} = \{1,\cdots, M \} $, respectively. For any given UAV $ m \in  \mathbf{M} $, the subset of IoT devices within its coverage is denoted by $N_m$. Furthermore, let $\mathbf{Q} $ represent the set of all processing nodes, which includes the other UAVs and the MEC server but excludes UAV $m$. This set is defined as $\mathbf{Q} = (\mathbf{M} \setminus \{m\}) \cup \{s\} = \{1,\cdots, Q\} $.

Let us adopt a priority-driven approach to task management, where each task is characterized by its data size and priority level. Specifically, $T_{n,k}$ represents task $k$ on the IoT device $n$. Therefore, for this task, the data size and priority level are denoted by $D_{n,k}$ and $p_{n,k}$, respectively. In our task offloading strategy, tasks are evaluated based on their designated processing location, either locally on a UAV or remotely on an edge node, reflecting the coverage of the IoT devices by its associated UAV. Local processing, indicated by binary variable $x_{n,k,m}$, occurs when a task $k$ from IoT device $n$ is processed on UAV $m$, utilizing the onboard resources of the UAV. Alternatively, edge processing, denoted by $y_{n,k,q}$, involves offloading tasks to an edge node $q$, leveraging other computational power. This strategy ensures that the processing location of each task is optimally chosen, balancing immediate processing needs against computational demands. Building upon these foundational definitions, we will next employ them to articulate both the communication and computational models of our system.

\subsection{Communciation Model}
This configuration includes two transmission stages: the first hop, termed N2M transmission, involves an uplink from IoT device $n$ to legitimate UAV $m$; the second hop, termed M2Q transmission, involves an uplink from UAV $m$ to the target node $q$. Further details regarding the features and parameters of these transmissions will be provided in the subsequent subsections.

\subsubsection{N2M Transmission}
We consider both LoS and Non-LoS (NLoS) propagation scenarios in the N2M transmission. The probability of LoS between an IoT device $n$ and UAV $m$ is determined as follows \cite{bao2021secrecy}:
\begin{align}
        p^{\text{LoS}}_{n,m}= \frac{1}{1+a \cdot \exp [-b(\theta_{n,m}-a)]},
\end{align}
where $a$ and $b$ represent environment-specific constants, and $\theta_{n,m}$ represents the elevation angle between IoT device $n$ and UAV $m$. The average path loss for this transmission is expressed by \cite{mozaffari2017mobile}:
\begin{align}
    L_{n,m}=  [\eta_1 p^{\text{LoS}}_{n,m} + \eta_2 (1- p^{\text{LoS}}_{n,m}) ] (K_0 d_{n,m})^\iota,
\end{align}
where $\eta_1$ and $\eta_2$ represent the excessive path loss coefficients under LoS and NLoS conditions, respectively. Here, $K_0= 4 \pi f_c/c$ incorporates the carrier frequency $f_c$ and the speed of light $c$. Moreover, $\iota$ denotes the path loss exponent, while $d_{n,m}$ refers to the distance between IoT device $n$ and UAV $m$. Consequently, the resulting average channel gain between IoT device $n$ and UAV $m$ can be experessed as $h_{n,m}= 1/L_{n,m}$. In order to mitigate potential interference and enable effective communication, orthogonal frequency-division multiple Access (OFDMA) is employed, allowing multiple access without interference. As a result, the transmission rate between an IoT device $n$ and UAV $m$ can be calculated as
\begin{align} \label{trasmrateiottouav}
    R_{n,m}=\frac{B}{N_m} \log_2 \left(1+\frac{p_n h_{n,m}}{N_0} \right),
\end{align}
with $B$ representing the total available bandwidth, $p_n$ representing the transmit power of IoT device $n$, and $N_0$ representing the noise power. 

We assume that Eve's channel experiences the path-loss fading effects as well. Thus, the average path loss from IoT device $n$ to Eve can be given by $L_{n,E}= [\eta_1 p^{\text{LoS}}_{n,E} + \eta_2 (1-p^{\text{LoS}}_{n,E}) ]  (K_0 d_{n,E})^\iota$.  Here, $p^{\text{LoS}}_{n,E}$ indicates the probability of LoS between IoT device $n$ and Eve, while $d_{n,E}$ denotes the distance between them.
Then, the average channel gain between IoT device $n$ and Eve $E$ is calculated as $h_{n,E} = 1/ L_{n,E}$. Consequently, the transmission rate between IoT device $n$ and Eve $E$ can be expressed by
\begin{align} \label{trasmrateiottoeve}
    R_{n,E}= \frac{B}{N_m} \log_2 \left(1+\frac{p_n h_{n,E}}{N_0} \right).
\end{align}
The corresponding secrecy rate can be expressed as the difference in transmission rates between the IoT device to legitimate UAV channel and the IoT device to Eve channel, given by \cite{leung1978gaussian}:
\begin{align} \label{secrecyrateiottouav}
    R_{n,s} = \max[ R_{n,m} - R_{n,E}, 0].
\end{align}
Considering task $k$ on the IoT device $n$, the transmission time required to securely send task $k$ from IoT device $n$ to UAV $m$ can be represented as shown below:
\begin{align} \label{trastimeiottouav}
      t_{n,s}= \frac{D_{n,k}}{R_{n,s}}.
\end{align}
To illuminate the trade-offs between energy efficiency and secure communication within IoT environments, the energy consumption resulting from transmitting task $k$ from IoT device $n$ to UAV $m$ during this period is determined by:
\begin{align} \label{energyiottouav}
      E_{n,s}=p_n t_{n,s}.
\end{align}
\subsubsection{M2Q Transmission}
UAV $m$ has the option to assign task $k$ to a specific target node within the set $q \in \mathbf{Q}$. This target node can either be another legitimate UAV or a MEC server, depending on the fulfillment of $y_{n,k,q} = 1$. In the M2Q transmission scenario, we only consider LoS propagation. Additionally, the presence of Eve can potentially compromise the secrecy integrity of the second-hop communication, warranting the need to account for its impact. As a result, the transmission rates from UAV $m$ to both the target node $q$ and the eavesdropper Eve $E$ can be expressed by:
\begin{align} \label{tramsrateuavtotarget}
      R_{m,q}=\frac{B}{N_q} \log_2 \left(1+\frac{p_m h_{m,q}}{N_0} \right),
\end{align}
and
\begin{align}  \label{tramsrateuavtoeve}
     R_{m,E}= \frac{B}{N_q} \log_2 \left(1+\frac{p_m h_{m,E}}{N_0} \right),
\end{align}
respectively, where $N_q$ denotes the number of nodes within the target set $q$, $p_m$ is the transmission power of UAV $m$, $h_{m,q}$ and $h_{m,E}$ are the channel gains to target node $q$ and eavesdropper Eve $E$ respectively, with $N_0$ being the noise power. The channel gains, $h_{m,q}$ and $h_{m,E}$, are determined by $h_{m,q} = 1/ [\eta_1 p^{\text{LoS}}_{m,q}  (K_0 d_{m,q})^\iota]$ and $h_{m,E} = 1/ [\eta_1 p^{\text{LoS}}_{m,E}  (K_0 d_{m,E})^\iota]$, respectively, where $d_{m,q}$ is the distance between UAV $m$ and target node $q$, and $d_{m,E}$ is the distance between UAV $m$ and Eve $E$.
Furthermore, for task $k$ hosted on UAV $m$, the corresponding secrecy rate, transmission time, and energy consumption for secure communication from UAV $m$ to the target node $q$ can be calculated as:
\begin{align} \label{secrecyrateuavtoq}
    R_{m,s} = \max[R_{m,q}-R_{m,E},0],
\end{align}
\begin{align} \label{tramstimeuavtoq}
    t_{m,s}= \frac{D_{n,k}}{R_{m,s}}, 
\end{align}
and
\begin{align} \label{energyuavtoq}
    E_{m,s}=p_m t_{m,s},
\end{align}
respectively. 

\subsection{Computational Model}
Considering the flexibility of processing task $k$,  either locally or offloading it for remote execution, we provide a detailed analysis of the computation time and energy consumption associated with both local and edge processing.

\subsubsection{Local Computing Model}
For UAV $m$, the feasibility of locally executing task $k$ is indicated by $x_{n,k,m} = 1$. As a result, the local computation time  is defined as:
\begin{align} \label{localt}
    t_{m,k}^{\text{loc}} = \frac{n_{k}^{\text{cpu}}}{f_m^{\text{loc}}},
\end{align}
where $n_{k}^{\text{cpu}}$ represents the number of CPU cycles necessary to process task $k$, and $f_m^{\text{loc}}$ specifies the CPU frequency dedicated to task $k$ by UAV $m$. The local energy consumption is calculated as follows:
\begin{align} \label{locale}
    E_{\text{loc}} = \kappa_{\text{loc}} (f_m^{\text{loc}})^2 n_{k}^{\text{cpu}},
\end{align}
with $\kappa_{\text{loc}}$ being the energy conversion coefficient for task $k$ that depends on the CPU's architectural design.

\subsubsection{Edge Computing Model}
If task $k$ is offloaded to a target node $q$ ($y_{n,k,q} = 1$), the edge computation time can be determined as:
\begin{align} \label{edget}
      t_{q,k}^{\text{edg}} = \frac{n_{k}^{\text{cpu}}}{f_q^{\text{edg}}},
\end{align}
where $f_q^{\text{edg}}$ represents the CPU frequency allocated for task $k$ by target node $q$. The calculation for the edge energy consumption is as follows:
\begin{align} \label{edgee}
      E_{\text{edg}} = \kappa_{\text{edg}} (f_q^{\text{edg}})^2 n_{k}^{\text{cpu}},
\end{align}
where $\kappa_{\text{edg}}$ denotes the energy conversion coefficient for task $k$, which is a factor determined by the CPU specifications of the target node. 

Therefore, the total delay and energy consumption for task $k$ from IoT device $n$ when assigned to UAV $m$ can be accurately summarized as follows:
\begin{align}
    T_{n,k,m} &= x_{n,k,m} (t_{n,s} + t_a +t_{m,k}^{\text{loc}})  \nonumber\\
    &+ y_{n,k,q} (t_{n,s} + t_{m,q} + t_a + t_{q,k}^{\text{edg}}),
\end{align}
and
\begin{align}
    E_{n,k,m} = x_{n,k,m} ( E_{n,s} + E_{\text{loc}}) + y_{n,k,q} ( E_{m,s} + E_{\text{edg}}),
\end{align}
respectively, where $t_a$ represents the decision-making time by UAV $m$ or target node $q$ to determine whether to process locally or offload remotely.

\subsection{Problem Formulation}
To enhance the efficiency and security of the transmission system, our goal is to minimize the system cost by optimizing task allocation and processing strategies. This objective includes reducing both delay and energy consumption, while ensuring secure communications. This can be achieved by prioritizing tasks based on their urgency and importance. Mathematically, we can formulate this optimization objective as follows:
\begin{flalign} \label{optplm}
\textbf{P}_{1}: \min_{\{x_{n,k,m}, y_{n,k,q}\}} &   \sum_{m=1}^{M} \sum_{n=1}^{N_m} \sum_{k \in K} \left(  p_{n,k} [\alpha  T_{n,k,m} + \beta  E_{n,k,m} ] \right) \text{, } \\
{\rm{s}}{\rm{.t}}{\rm{. }} \,\, 
& C1: T_{n,k,m}  \leq t_{\text{th}}, \quad \forall n, k, m \nonumber\\
& C2: R_{n,s} \geq R_{\text{min}}, \quad \forall n, k \nonumber\\
& C3: R_{m,s} \geq R_{\text{min}}, \quad \forall m, q, k\nonumber\\
& C4: E_{n,s}  \leq E_{n}^{\text{max-trans}}, \quad \forall n \nonumber\\
& C5: E_{m,q}  \leq E_{m}^{\text{max-trans}}, \quad \forall m, q \nonumber\\
& C6: E_{\text{loc}}  \leq E_{n}^{\text{max-proc}}, \quad \forall m,k \nonumber\\
& C7: E_{\text{edg}}  \leq E_{m}^{\text{max-proc}}, \quad \forall q,k \nonumber\\
& C8: x_{n,k,m} + y_{n,k,q} = 1, \quad \forall n, k, m \nonumber\\
& C9: \sum_{n=1}^{N_m} \sum_{k \in K } D_{n,k} \cdot x_{n,k,m} \leq C_m, \quad \forall m , \nonumber
\end{flalign}
where $\alpha$ and $\beta$ are weighting factors used to calibrate the significance of reducing delay and energy consumption, respectively. Constraint $C1$ addresses the limitations on delay for both local and edge processing, ensuring that the combined processing and transmission time for each task remains below a threshold $t_\text{th}$. Constraints $C2$ and $C3$ enforce the requirements fore secrecy rate in secure communication across both transmission hops. Constraints $C4$ through $C7$ set upper bounds on energy usage for both transmission and processing activities. $C8$ stipulates that each task must be designated either for local processing on a UAV or for offloading. Finally, $C9$ imposes capacity restrictions on each UAV, ensuring that the total data size of tasks processed does not surpass the UAV's capability $C_m$. 

The goal of this optimization function $\textbf{P}_{1}$ is to minimize the overall system cost while ensuring secure communications, by adjusting the location of task processing, i.e., deciding which tasks should be processed locally on UAVs and which should be offloaded to edge servers. Given the involvement of integer decision variables indicating whether a task is processed on a specific UAV or offloaded and non-linear relationships, e.g., energy consumption proportional to the square of processing frequency, this forms a MINLP with challenging solution complexity.

\section{Multi-agent DRL based on action mask}
To efficiently address this issue, we transformed the problem into a markov decision process (MDP) framework. We propose a multi-agent DRL-based method that integrates the DDQN with AM and the Knapsack algorithm. This novel approach combines the strengths of reinforcement learning for dynamic decision-making and classical optimization to effectively handle resource allocation under constraints effectively. In our multi-agent DRL system, UAVs are defined as individual agents. Thus, for each UAV $m$, the MDP framework defines the following spaces:

\subsection{State Space}
In a multi-agent DRL framework, specifically designed for a UAV-assisted smart farm network, the state space encompasses all the requisite information for each UAV to make informed decisions regarding task offloading and processing. In every time slot $t$, the state space for UAV $m$ can be expressed as follows:
\begin{align}
  s_t^m =[ B_m, C_m, \{T_k[D,p]\}_{k=1}^K, \{R_k\}_{k=1}^K ],
\end{align}
where $B_m$ is the battery power of UAV $m$, where a critical parameter determining the UAV's ability to process and offload tasks, $C_m$ represents the computational capacity of UAV $m$, indicating the UAV's current available computational resources. $\{T_k[D,p]\}_{k=1}^K$ encapsulates the set of tasks, ranging from 1 to $K$, where each task $k$ characterized by its data size $D$ and priority $p$, reflecting the tasks' requirements and importance. Finally, $\{R_k\}_{k=1}^K$ indicates the outcomes from applying the Knapsack algorithm to each task $k$, revealing whether the task has been selected for local processing on the UAV. In this DRL paradigm, the UAVs' policy networks leverage the state space as input, facilitating the generation of optimal actions that govern task offloading and processing strategies. This approach ensures that decisions are made with a comprehensive understanding of each UAV's operational capabilities and the task-specific demands, thereby optimizing the network's overall efficiency and responsiveness.

\subsection{Action Mask}
In a complex environment such as a UAV-assisted smart farm network, an AM emerges as a crucial mechanism. It ensures that each agent (UAV) considers only feasible actions at any given time, based on its current state and operational constraints. The AM for each task $k$ and UAV $m$, in each time slot $t$, can be defined as:
\begin{align}
   M^{m}_{t} = 
   \begin{cases} 
      1 & \text{if action $ a_{m,k} $ is feasibility given $s_t^m$},\\
      0 & \text{otherwise},
   \end{cases} 
\end{align}
where feasibility can be determined based on the following factors: battery power of UAV $m$, $B_m$, which ensures the UAV has sufficient energy to complete the task, and computational capacity of UAV $m$, $C_m$, which checks if the UAV has enough processing power to handle the task locally. Before a UAV decides on an action for a task, it consults the AM. Functioning as a selective filter, the AM permits only those actions deemed feasible within the current operational state and existing constraints. Consequently, a UAV proceeds to undertake only those actions that have been validated by the AM.

\subsection{Action Space}
In the time slot $t$, the action space for UAV $m$ can be represented by
\begin{align}
    a_{t}^m = [ a_{m,1}, \cdots, a_{m,k},\cdots, a_{m,K} ],
\end{align}
where $a_{m,k} \in \{0,1\}$ represents the binary decision associated with each task $k$. In a scenario where UAV $m$ is presented with $K$ tasks, each capable of being processed locally or offloaded, the action space expands to include $2^K$ different action combinations. Each of these combinations represents a unique set of decisions across all tasks.

\subsection{Reward Space}
The optimization objective of the reward function within a secure UAV-assisted smart farm network is intricately designed to reduce system costs while simultaneously guiding the agent towards the maximization of long-term rewards. This dual-faceted goal ensures that immediate operational efficiencies are achieved without compromising the strategic pursuit of future benefits. Aligned with the optimization function $\textbf{P}_{1}$, the global reward function in time slot $t$ is refined as follows:
\begin{align}
       R_{t} = - \sum_{m=1}^{M}\sum_{n=1}^{N_m} \sum_{k \in K} \left( p_{n,k} [\alpha T_{n,k,m} + \beta E_{n,k,m} ] \right). 
\end{align}

\subsection{Learning Alogorithm}

\begin{figure}[t!]
\centering
\includegraphics[height=7.8cm,width=8.8cm]{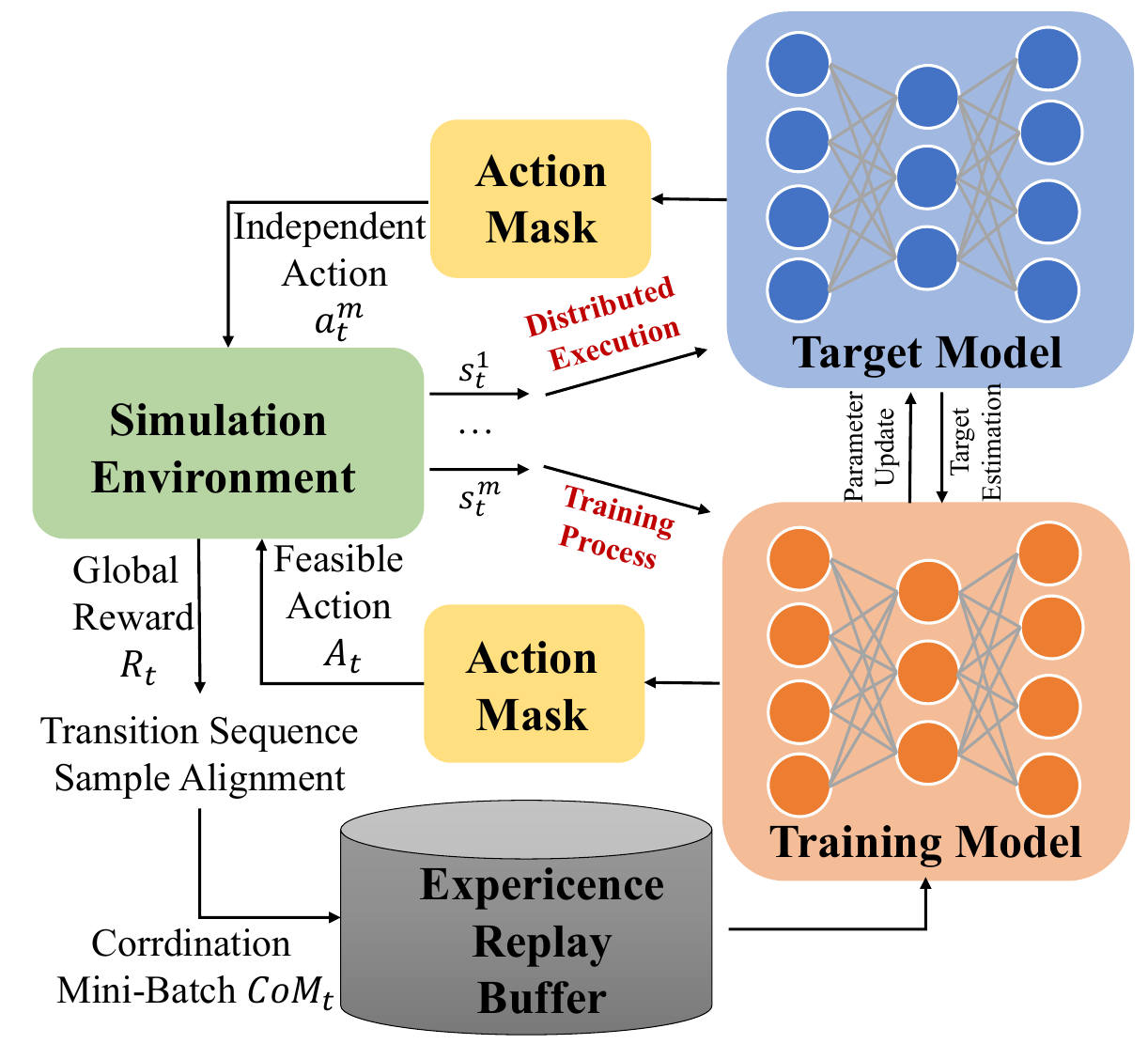}
\caption{Framework of our proposed multi-agent DRL based on AM.}
\label{fig_alg_mdl}
\end{figure}

In our multi-agent DRL framework designed for UAV-assisted smart farm networks, we implement a training process coupled with a distributed execution stage. This dual-phase approach optimizes collective learning and individual operational usefulness, ensuring that each UAV not only contributes to the overarching network goals but also navigates its immediate environment with enhanced autonomy and precision, as illustrated in Fig. \ref{fig_alg_mdl}.
\subsubsection{Training Process}
The training process is a meticulous procedure designed
to cultivate cooperative behavior among UAVs in order to attain optimal system performance. This procedure encompasses a number of crucial stages:

\begin{itemize}
    \item \textbf{Experience Collection}: During this phase, each UAV agent interacts with the environment by performing actions based on its current policy and observing the outcomes. These interactions are encapsulated as experiences, which are defined for each time step $t$ and each UAV $m$ as:
\begin{align}
e_t^m = (s_t^m, a_t^m, R_t, s_{t+1}^m, M_t^m),
\end{align}
where $s_{t+1}^m$ represents the next state resulting from the action.
   \item \textbf{Experience Storage}: These experiences are then stored in an experience replay buffer ($\mathcal{D}$), which aggregates the data from all UAVs:
\begin{align}
 \mathcal{D} \leftarrow \mathcal{D} \cup \{e_t^m\} \quad \forall m \in M.
\end{align}
This pool serves as a comprehensive database of the intercations between the the multi-agent system and the environment, enabling the learning algorithm to access a diverse set of experiences.
   \item \textbf{Coordination Mini-Batch Sampling}: For effective training, experiences are sampled from $\mathcal{D}$ in the form of Coordination Mini-Batches (CoM). Each CoM sample contains synchronized experiences from all UAVs within the same time slot, ensuring that the training process considers the agents' actions in conjunction:
\begin{align}
   \text{CoM}_t = \{e_t^1, e_t^2, \ldots, e_t^M\} .
\end{align}
This promotes implicit cooperation by aligning the agents' learning objectives with the network's overall goals.
Coordination Mini-Batch Sampling
   \item \textbf{Network Updates}: The training and target networks of the DDQN algorithm are updated using the sampled experiences. Therefore, the loss function for updating the training network can be defined as:
\begin{align} \label{loss}
  \mathbb{E}_{(s, a, R, s') \sim {\mathcal{D}}_{\text{CoM}_t}  } &\bigg[\bigg(R + \gamma \max_{a'}Q(s', a'; \phi')  \nonumber\\
  & - Q(s, a; \phi)\bigg)^2 \bigg] , 
\end{align}
where $\phi$ and $\phi'$ represent the parameters of the training and target networks, respectively, while $\gamma$ stands for the discount factor.
 \item \textbf{Periodic Network Synchronization}: The parameters of the target network ($\phi'$) are updated with those from the training network ($\phi$ ) at intervals, ensuring a stable progression of learning.
\end{itemize}
\subsubsection{Distributed Execution Stage}
Following the centralized training, UAVs transition to the distributed execution stage. In this stage, each agent independently applies the learned policy based on localized observations and decision-making. This can be given by:
\begin{align}
    a_{t}^m = \arg\max_{a \in A_{t}^m, M_{t}^m (a) = 1} Q(s_{t}^m, a; \phi),
\end{align}
where $ a_{t}^m$ represents the action selected by UAV $m$ at time $t$, maximized over the feasible action space $A_{t}^m$ as determined by the action mask $M_{t}^m$, ensuring adherence to operational constraints and local context. 

The details of our proposed algorithm for a UAV-assisted smart farm network employing multi-agent DRL are comprehensively outlined in Algorithm \ref{alg_1}. This structured presentation encapsulates the sequence of steps from initial setup through centralized training to the transition into distributed execution, ensuring clarity in understanding both the theoretical framework and practical implementation.

\begin{algorithm} [t!]
\caption{Multi-agent DRL based on AM for secure UAV-assisted Smart Farm Networks}
\begin{algorithmic}[1]  \label{alg_1}
\STATE Initialize training network parameters $\phi$ and target network parameters $\phi'$ for each UAV
\STATE Initialize shared experience replay pool $\mathcal{D}$
\STATE Set AM criteria based on operational constraints
\FOR{each episode}
    \STATE Reset environment and obtain initial state $s_0^m$ for each UAV $m$
    \FOR{each time step $t$ within the episode}
        \FOR{each UAV $m$}
            \STATE Apply AM to determine feasible actions $A_t^m$ based on $s_t^m$
            \STATE Select action $a_t^m$ using $\epsilon$-greedy policy from $\phi$
            \STATE Execute $a_t^m$, observe reward $R_t$, next state $s_{t+1}^m$
            \STATE Update action mask $M_t^m$ for the next decision
            \STATE Store $(s_t^m, a_t^m, R_t, s_{t+1}^m, M_t^m)$ in $\mathcal{D}$
        \ENDFOR
        \IF{it's time to update}
            \STATE Sample CoM from $\mathcal{D}$
            \STATE Calculate target values using $\phi'$ and observed $R_t$
            \STATE Update $\phi$ by minimizing loss utilizing Eq. (\ref{loss}) between predicted Q-values and target values
            \STATE Every $C$ steps, update $\phi'$ with weights from $\phi$
        \ENDIF
    \ENDFOR
\ENDFOR
\STATE \textbf{Transition to Distributed Execution:}
\FOR{each UAV $m$ in execution mode}
    \STATE Independently select actions based on trained policy and local observations using AM
    \STATE Execute selected actions and adapt to outcomes
\ENDFOR
\end{algorithmic}
\end{algorithm}

\section{SIMULATION RESULTS}
In this section, we verify and discuss the proposed multi-agent DRL framework, which incorporates AM, through simulations designed for secure UAV-assisted smart farm networks. Detailed descriptions of the environment settings and model parameters for these simulations are provided in Table \ref{tab_1}. For our system model, we assume that IoT devices, UAVs, and Eve are initially positioned randomly within a three-dimensional space measuring 100m $\times$ 100m $\times$ 100m. The DRL algorithms are implemented using the TensorFlow framework, a choice that offers advanced optimization tools and flexibility in designing complex neural network architectures. The associated hyperparameters are detailed in Table \ref{tab_2}. We employ fully connected neural networks comprising three layers to facilitate the learning process. The training regimen is configured with a batch size of 300, utilizing an experience replay pool size of $10^4$. The entire training process spans 1,000 episodes, ensuring comprehensive learning and robust model evaluation.

\begin{table}[t!] 
\renewcommand{\arraystretch}{1.35}
\centering  
\caption{System Parameters} \label{tab_1}
\begin{tabular}{m{5.5cm}|m{2.4cm}}

\hline
\hline
Parameter & Value \\
	   
\hline
Number of IoT devices, $N$ & $[3,7,10]$ \\
\cdashline{1-2}[0.8pt/1pt]
Number of task, $K$ & 3\\
\cdashline{1-2}[0.8pt/1pt]
Number of UAV, $M$ & 4 \\
\cdashline{1-2}[0.8pt/1pt]
System bandwidth, $B$ & 20 MHz \\
\cdashline{1-2}[0.8pt/1pt]
Noise power sepctral density, $N_0$ & -96 dBm  \\
\cdashline{1-2}[0.8pt/1pt]
Space area & [100,100,100]\\
\cdashline{1-2}[0.8pt/1pt]
Spatial distribution of IoT devices/UAV/Eve & Uniform\\
\cdashline{1-2}[0.8pt/1pt]
Carrier frequency, $f_c$ & 2.4 GHz \\
\cdashline{1-2}[0.8pt/1pt]
Path loss exponent, $\iota$ & 3.0 \\
\cdashline{1-2}[0.8pt/1pt]
Transmit power of IoT devices, $P_n$ & 15 dBm \\
\cdashline{1-2}[0.8pt/1pt]
Transmit power of UAV, $P_m$ & 23 dBm \\
\cdashline{1-2}[0.8pt/1pt]
Air-to-ground channel model & Rural area with $[a=11.25, b=0.06]$\\
\cdashline{1-2}[0.8pt/1pt]   
Data size of the task, $D_{n,k}$&  Gaussian distribution with $\mathcal{N} (1,0.1) $ MB  \\
\cdashline{1-2}[0.8pt/1pt]   
Priority level of the task, $p_{n,k}$&  $[0.3,0.6,0.9]$  \\
\cdashline{1-2}[0.8pt/1pt]  
Number of CPU-cyle requirements for task, $n_k^{\text{cpu}}$& Gaussian distribution with $\mathcal{N} (100,10) $ Megacycles  \\
\cdashline{1-2}[0.8pt/1pt] 
Battery power of UAV, $B_m$ & 3e4 J \\
\cdashline{1-2}[0.8pt/1pt]  
Computational capacity of UAV, $C_m$ & 3 MB \\  
\cdashline{1-2}[0.8pt/1pt]
CPU refquency for UAV  & 100 MHz \\
\cdashline{1-2}[0.8pt/1pt]
CPU refquency for MEC server & 500 MHz \\
\cdashline{1-2}[0.8pt/1pt]
Energy conversion coefficient for UAV  & 1e-16 \\
\cdashline{1-2}[0.8pt/1pt]
Energy conversion coefficient for MEC server & 1e-22 
\\

\hline
\hline
\end{tabular}
\end{table}

\begin{table}[t!] 
\renewcommand{\arraystretch}{1.35}
\centering  
\caption{Hpyer parameters of DDQN model} \label{tab_2}
\begin{tabular}{m{5.5cm}|m{2.4cm}}

\hline
\hline
Parameter & Value \\
	   
\hline
Number of neurons in each layer
 & (32,64,128) \\
\cdashline{1-2}[0.8pt/1pt]
Activation function &
ReLU \\ 
\cdashline{1-2}[0.8pt/1pt]
Optimizer &
Adam \\
\cdashline{1-2}[0.8pt/1pt]
Learning rate &
1e-4 \\
\cdashline{1-2}[0.8pt/1pt]
Discount rate &
0.9 \\
\cdashline{1-2}[0.8pt/1pt]
Batch size &
300 \\
\cdashline{1-2}[0.8pt/1pt]
Experience Replay Pool &
10000 \\
\cdashline{1-2}[0.8pt/1pt]
Episode &
1000 \\
\hline
\hline
\end{tabular}
\end{table}

We incorporate two baseline strategies to assess the performance of our proposed DDQN with AM algorithm. Firstly, we employ a random strategy as a baseline, which selects actions randomly within the DRL framework, disregarding historical data or learning mechanisms. This serves as a control condition that reflects decision-making without strategic guidance. Secondly, we utilize a DDQN without AM as another baseline strategy. In this setup, the DDQN algorithm makes decisions based on traditional reinforcement learning techniques, without the benefit of targeted exploration provided by AM. 

\begin{figure}[t!]
\centering
\includegraphics[height=6cm,width=7.8cm]{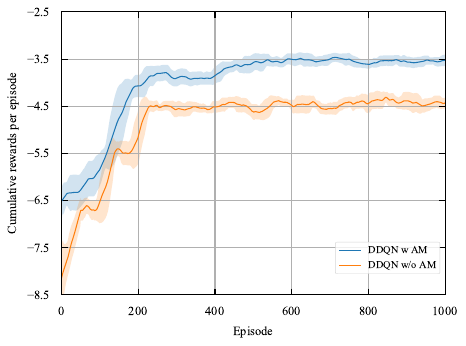}
\caption{Reward convergence versus episodes for different algorithms, with $N=10$. }
\label{fig_reward_algri}
\end{figure}

Fig. \ref{fig_reward_algri} illustrates the convergence of rewards over training episodes for two different algorithms: DDQN without AM, and our proposed DDQN with AM. As the number of training episodes increases, the cumulative rewards for our proposed DDQN with AM initially rise sharply before reaching a stable state, indicating successful convergence of our algorithm. A comparative analysis  reveals that our proposed DDQN with AM not only consistently outperforms the others in terms of average rewards but also demonstrates a more rapid initial increase in rewards, followed by stable convergence at a higher reward baseline. Additionally, AM enhances the exploration process in reinforcement learning by filtering out inappropriate actions, which helps in acquiring higher-quality samples. This selective exploration contributes to the superior performance of DDQN with AM over DDQN without AM, as depicted in the figure. Consequently, our proposed DDQN with AM significantly enhances the stability and overall performance of DRL model training.

\begin{figure}[t!]
\centering

\begin{subfigure}[t]{0.5\textwidth}  
    \centering
    \includegraphics[height=5.8cm,width=7.8cm]{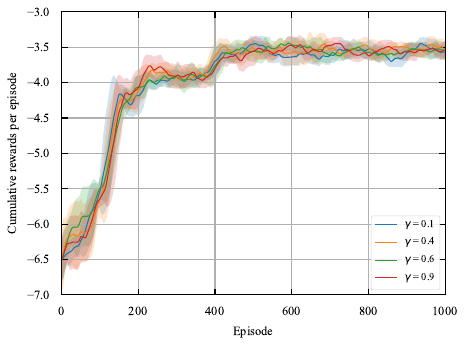}  
    \caption{}
    \label{fig:sub1}  
\end{subfigure}
\begin{subfigure}[t]{0.5\textwidth}  
    \centering
    \includegraphics[height=5.8cm,width=7.8cm]{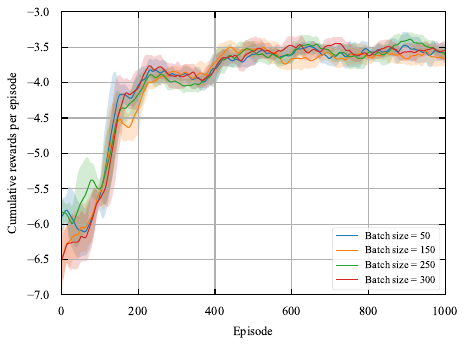}  
   \caption{}
    \label{fig:sub2}  
\end{subfigure}

\caption{Cumulative reward of DDQN with AM under differnt hyperparameters: (a) discount rate $\gamma$ (top), (b) batch size (bottom), $N=10$.}  
\label{fig:verticalsubfigures}  
\end{figure}

Fig. \ref{fig:verticalsubfigures} illustrates the cumulative rewards per episode for the DDQN with AM across varying discount rates and batch sizes. The analysis of the data reveals that modifications to either the discount rate or the batch size do not have a significant impact on the model's convergence stability. This observation is particularly noteworthy as it demonstrates that our proposed DDQN with AM algorithm effectively mitigates some of the common issues faced in DRL models, such as sensitivity to hyperparameters and unstable training phases. The model's robustness under varying hyperparameter settings underscores its strong adaptability, making it an advantageous choice for environments where hyperparameter tuning can be challenging.

\begin{figure}[t!]
\centering
\includegraphics[height=5.8cm,width=7.8cm]{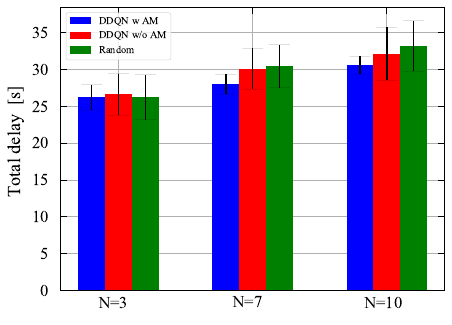}
\caption{Total delay versus the number of IoT devices for differnet algorithms.}
\label{fig_delay}
\end{figure}

\begin{figure}[t!]
\centering
\includegraphics[height=5.8cm,width=7.8cm]{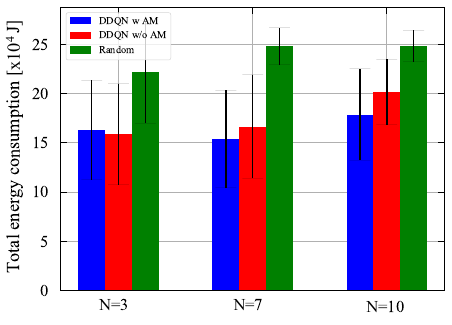}
\caption{Total energy consumption versus the number of IoT devices for differnet algorithms.}
\label{fig_energy}
\end{figure}

Figures \ref{fig_delay} and \ref{fig_energy} present a comparative analysis of network performance in terms of total delay and energy consumption, respectively, for three distinct algorithms: Random, DDQN without AM, and our proposed DDQN with AM. These metrics have been evaluated across varying numbers of IoT devices. As shown in figures, both total delay and energy consumption increase as the number of IoT devices increases, which aligns with the expected grownth in data processing and communication demands associated with larger device arrays. Additionally, the figures demonstrate the performance benefits of our proposed DDQN with AM become more prominent as the number of IoT devices increases. This improvement is especially significant in high IoT density enviroments, where the efficient management of larger and more complex action spaces by DDQN with AM becomes crucial. These environments highlight the robustness and enhanced efficiency of DDQN with AM, making it as a superior choice for scenarios that require high scalability and operational efficiency.

\section{Conclusion}
In this paper, we investigated priority-driven task offloading in secure UAV-assisted smart farm networks. We formulated an optimization problem aimed at minimizing the total delay and energy consumption while ensuring secure communications. This was achieved by prioritizing tasks based on their urgency and importance. To address this challenge, we proposed a DDQN with AM scheme, which incorporates both a training process and a distributed execution stage framework. Through simulations, we validated the convergence and superiority of our proposed scheme. Thus, the proposed method not only enhances of operational efficiency and sustainability but also offers practical solutions for real-world deployment in the next-generation network infrastructures, which can address critical performance and energy managment challenges in smart agriculture. Future work will explore the broader applicability of our approach in other domains, such as smart cites and industries, further contributing to sustainable and efficicent network operations.

\bibliographystyle{IEEEtran}
\bibliography{sdor}
\end{document}